\documentstyle[11pt,newpasp,twoside,epsf]{article}
\begin{document}
\title{Geodetic Precession and the Binary Pulsar B1913+16}
\author{A. Karastergiou, M. Kramer, N. Wex, A. von Hoensbroech}
\affil{
Max Planck Institut f\"ur Radioastronomie, Auf dem H\"ugel 69, 53121 Bonn, Germany
}
\begin{abstract}
A change of the component separation in the profiles of the binary pulsar PSR B1913+16
has been observed for the first time (Kramer 1998) as
expected by geodetic precession.
In this work we extend the previous work by accounting
for recent data from the Effelsberg 100-m telescope and Arecibo Observatory
and testing model predictions. We demonstrate how the new information will provide additional
information on the solutions of the system geometry.
\end{abstract}

\vspace{-3ex}

\section{Introduction}

The binary pulsar PSR B1913+16 exhibits
a measurable amount of geodetic precession due to a misalignment of its
spin axis and orbital angular momentum (relativistic spin orbit coupling) by an angle $\delta$
(see also Stairs et al., for geodetic precession in PSR B1534+12).
The angle between the line of sight and the spin axis, $\zeta$, changes according to:
\begin{displaymath}
\cos\zeta(t) = -\cos\delta\cos i
                  +\sin\delta\sin i \cos \Omega_{{\rm prec}} (t-T_0) \;,
\end{displaymath}
where $i$ is the orbital inclination of the binary system.
General Relativity (GR) predicts a rate of precession $\Omega_{{\rm prec}}=1.21$ deg yr$^{-1}$,
leading to a change in the pulse profile and polarization characteristics with time.
\\
Previous observations by Weisberg, Romani and Taylor (1989) detected
a change in the relative amplitude of the trailing and leading component, but no
change in the component separation. Also, there were no
changes in the position angle swing of the linear polarization noted (Cordes, Wasserman \&
Blaskiewicz 1990).

\section{Recent Observations Cast Some Light}

The evolution of the observed radio pulse profile of PSR B1913+16 would be
determined when the four parameters
$\alpha, \delta, \rho, T_{0}$ were specified for a given $i$, where
$\alpha$ is the angle between the magnetic pole and the spin axis,
$\delta$ is as mentioned above,
$\rho$ is the angular radius of the emission cone and
$T_{0}$ the epoch of precession.
\nopagebreak
Based on recent data from the Effelsberg 100-m telescope, Kramer (1998) first noticed a
change in the component separation $W$. This allows a solution region in parametric space
that respects GR and the
{\bf{\underline{hollow cone model} (HCM)}}.
\begin{displaymath}
 W(t) = 2\arccos\left[\frac{\cos\rho - \cos\alpha \cos\zeta(t)}
                             {\sin\alpha \sin\zeta(t)}\right] 
\end{displaymath}
Effelsberg polarization data has been used
in respect with the \underline{\bf rotating vector} \\ \underline{\bf model} {\bf(RVM)},
to further narrow down the solution region.
\begin{displaymath}
 \psi(t) = \arctan\left[\frac{\sin\alpha\sin(\phi-\phi_0)}
      {\cos\alpha \sin\zeta(t) - \sin\alpha\cos\zeta(t)}\right] \quad + const. 
\end{displaymath}

The inclination of the binary orbit with respect to the line-of-sight, $i$,
is known from timing observations modulo the ambiguity
$i\longrightarrow 180^\circ - i$. The possibilities are:
Case A:  $i= 47.2^\circ$,  Case B:  $i= 132.8^\circ$ 

\vspace{-2ex}

\section{The Solution Today}

\begin{figure}[h]
\centerline{
\epsfxsize=7cm
\epsfysize=7cm
\epsffile{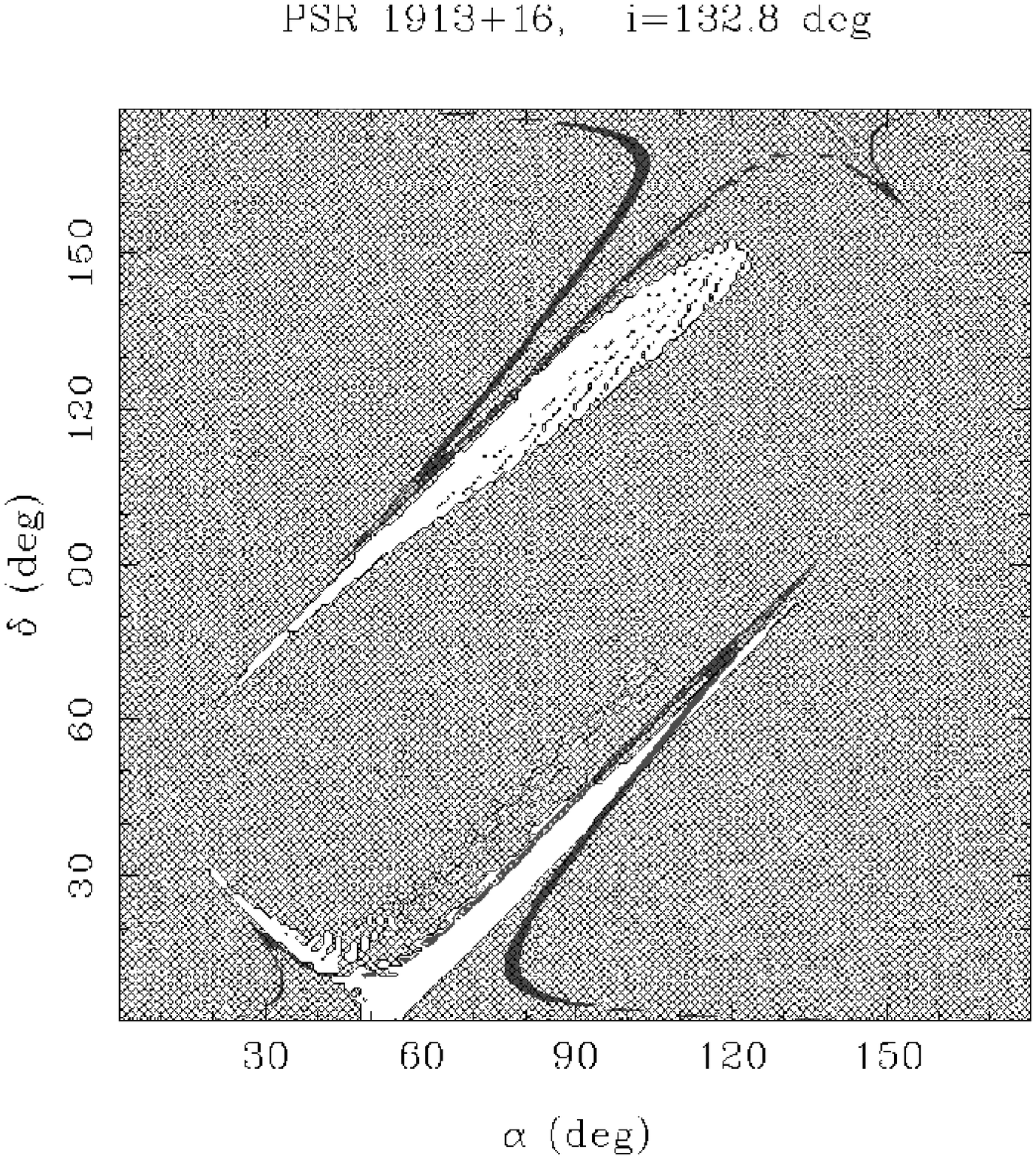}
\hspace{0.25cm}
\parbox[b]{6cm}{\sloppy
Figure 1. {\small The pulse component separation data (Effelsberg + Arecibo) in conjunction with the HCM
was used to produce a $\chi^{2}$ area plot indicating the $3\sigma$ confidence level
solutions for $\alpha$ and $\delta$, areas inside the surfaces traced by the black lines.
All possible solutions were then fed into the RVM to produce
position angle curves, which were compared to Effelsberg polarization data,
only allowing for solutions in the unshaded areas. So, the solution must be located on
a black surface
inside an unshaded region (for detailed explanation of this figure check out 
{\it http://www.mpifr-bonn.mpg.de/staff/akara/psr1913})
}}}
\end{figure}

The set of plausible solutions has been well limited, as shown in Fig 1.
The inclination of the binary orbit is most likely to be Case B rather than Case A (Wex, Kalogera, Kramer
2000). More precise use of the polarization data plus a molding of the
two models (HCM \& RVM) to specifically suit PSR B1913+16
(see also Weisberg \& Taylor in these proceedings)
could further limit the solutions and
enable a prediction on the future of this system.


\vspace{-2ex}

\begin{references}
\vspace{-2ex}
\reference
Cordes,~J.M., Wasserman,~I., \& Blaskiewicz,~M. 1990, \apj, 349, 546

\reference
Damour,~T., \& Ruffini,~R. 1974, Acad. Sci. Paris, 279, s\'erie A, 971

\reference
Damour,~T., \& Taylor,~J.H. 1992, Phys. Rev. D, 45, 1840

\reference
Hulse,~R.A., \& Taylor,~J.H. 1975, \apj, 195, L51

\reference
Kramer,~M. 1998, \apj, 509, 856

\reference
Taylor,~J.H. 1999, Proceedings of the XXXIV'th Rencontres de Moriond, to be published

\reference
Weisberg,~J.M., Romani,~R., \& Taylor,~J.H. 1989, \apj, 347, 1029

\reference
Wex,~N., Kalogera,~V., Kramer,~M. 2000, \apj, 528, 401

\end{references}
\end{document}